# A phase-space approach to non-stationary nonlinear systems

Vladimir L. Kalashnikov[1], and Sergey L. Cherkas[2]

[1] Institute of Photonics, Vienna University of Technology, Vienna, Austria
 (E-mail: vladimir.kalashnikov@tuwien.ac.at)
[2] Institute for Nuclear Problems, Belarus State University, Minsk, Belarus
 (E-mail: cherkas@inp.bsu.by)

**Abstract.** A phase-space formulation of non-stationary nonlinear dynamics including both Hamiltonian (e.g., quantum-cosmological) and dissipative (e.g., dissipative laser) systems reveals an unexpected affinity between seemly different branches of physics such as nonlinear dynamics far from equilibrium, statistical mechanics, thermodynamics, and quantum physics. One of the key insights is a clear distinction between the "vacuum" and "squeezed" states of a non-stationary system. For a dissipative system, the "squeezed state" (or the coherent "concentrate") mimics vacuum one and can be very attractable in praxis, in particular, for energy harvesting at the ultrashort time scales in a laser or "material laser" physics including quantum computing. The promising advantage of the phase-space formulation of the dissipative soliton dynamics is the possibility of direct calculation of statistical (including quantum) properties of coherent, partially-coherent, and non-coherent dissipative structure without numerically consuming statistic harvesting.

## 1 Introduction

The study of the dynamics of self-organized dissipative systems could bridge the alas different shores of our knowledge, and it has to be based on an understanding of a multiscale nature of underlying phenomena. Here, we shall try to demonstrate as the most general and, nevertheless, outwardly disjoined concepts can contribute productively to the study of nonlinear dynamics of nonequilibrium nonlinear systems. The keystone here is a phase-space formulation of a problem which reveals the intrinsic affinity between both classical and quantum Hamiltonian as well as non-Hamiltonian systems. Such affinity promises a breakthrough in the study and practical mastering of scalable coherent structures in the midst of noisy dissipative environment. The application area ranges from neurophysiology to quantum computing and high-energy laser physics.

———————

Here, we intend to illustrate the Weyl-Wigner-Moyal approach to the construction of the phase-space representation of seemly dissimilar systems ranging from quantum cosmology to ultrafast laser physics. The statistical mechanics and the theory of turbulence phenomena are the bearings in this enterprise.

There is a deep and physically relevant analogy between the evolutional laws for a mixed state of a quantum system (whether "closed" or "open") and the statistical mechanics. The Hamiltonian formulation of classical mechanics reveals this elegant and genuine kinship.

Let us remind the von Neumann law for the density matrix $\rho \equiv \sum_i P_i |\psi_i\rangle\langle\psi_i|$ evolution [1]:

$$\frac{\partial \rho(t)}{\partial t} = \frac{i}{\hbar}\left[\rho(t), H(t)\right], \qquad (1.1)$$

where $H(t)$ is the time-dependent Hamiltonian of a system, including, in the general case, the "environment" ("basin") and the interactional parts ($[*,*]$ denotes a commutator). This equation is a direct analog to the famous Liouville equation for the evolution of a phase-space distribution function $\rho$ in the statistical mechanics:

$$\frac{\partial \rho}{\partial t} = -\{\rho(t), H(t)\}, \qquad (1.2)$$

($\{*,*\}$ denotes the Poisson bracket) and to the law of evolution of a dynamical variable $A(t)$ within the frameworks of Hamiltonian formulation of classical mechanics:

$$\frac{dA(t)}{dt} = \{H(t), A(t)\}. \qquad (1.3)$$

However, the conceptual difference is that the phase space in the quantum mechanics is the operator space, and these operators can be noncommitting in the general case. A study of this space is a mathematically challenging issue, and, we face the interpretation challenges additionally. The instance of such problem, which is relevant to our work, is the practically useful definition of *vacuum state of a time-dependent quantum system* (e.g., the time-dependent quantum oscillator) and its distinguish from a so-called "squeezed state." The classical definition implying the vacuum state $|0\rangle$ as a "zero space" of annihilation operator $\hat{a}|0\rangle = 0$ is not practically useful in many cases. The important insight of the Hamiltonian minimization $\langle 0|\hat{H}|0\rangle$ is closely related to the situations, which will be considered below. At last, the asymptotical "uncertainty minimization" criterium $\left\langle |\hat{p} - \langle\hat{p}\rangle|^2 |\hat{x} - \langle\hat{x}\rangle|^2 \right\rangle = \frac{1}{4} + \sigma^2 = \frac{1}{4}\left(1 + \langle 0|\hat{x}\hat{p} + \hat{p}\hat{x}|0\rangle^2\right)$ [2] is relevant to important both quantum and classical problems.

## 2 A time-dependent (driven) quantum oscillator

An issue of time-dependent (driven) oscillator arises naturally in some fields of the theoretical physics. In particular, it has an application in cosmology and astrophysics, where the scalar, fermion, gravitational, and other quantum fields evolve in an expanding Universe. Nevertheless, the definition of the ground (vacuum) state remains to be obscure. It would be desirable to define vacuum state without appealing to the adiabatic series or analytical solution that can be impossible in praxis. This issue is addressed in the suggested method, which allows finding the true vacuum state numerically if such a state exists.

Let us remind the problem in more detail. The Hamiltonian of a time-dependent oscillator has the following form:

$$H = \frac{1}{2}\left(\dot{x}^2 + \omega^2(t)x^2\right). \quad (2.1)$$

The standard commutation relations for the momentum and coordinate operators are:

$$\left[\hat{p}(t), \hat{x}(t)\right] = \left[\hat{\dot{x}}(t), \hat{x}(t)\right] = -i. \quad (2.2)$$

The mean value of the kinetic and potential energies difference is expressed as

$$\langle 0|\frac{1}{2}\hat{p}^2(t) - \frac{1}{2}\omega(t)\hat{x}^2(t)|0\rangle = \dot{\sigma}(t). \quad (2.3)$$

Here

$$\sigma = \frac{1}{2}\langle 0|\hat{x}(t)\hat{p}(t) + \hat{p}(t)\hat{x}(t)|0\rangle \quad (2.4)$$

a sense of the additional uncertainty arising in the Heisenberg uncertainty relation:

$$\left\langle |\hat{p} - \langle\hat{p}\rangle|^2 |\hat{x} - \langle\hat{x}\rangle|^2 \right\rangle > \frac{1}{4} + \sigma^2 \quad (2.5)$$

and $\langle\ |, |\ \rangle$ are arbitrary states. For a family of the squeezed states, including a true vacuum, the inequality (2.5) becomes equality.

The straightforward computation shows that $\sigma$ satisfies the nonlinear equation

$$\left(4\sigma\omega^2 + \ddot{\sigma}\right)\left(4\sigma\omega^3 + \ddot{\sigma}\omega - 2\dot{\sigma}\dot{\omega}\right) - \omega\dot{\omega}^2\left(4\sigma^2 + 1\right) = 0, \quad (2.6)$$

for the states belonging to a family of the squeezed vacuum states including the true vacuum ones. Thus, one has the nonlinear equation (2.6) for choosing the true vacuum state from a family of the squeezed states. The nonlinearity in (2.6) arises from (2.2). We suggest that a true vacuum state corresponds to the monotonic time-dependence of $\sigma(t)$.

Since the criterium of a monotonic behavior of the $\sigma(t)$-function within a time interval $\{t_1, t_2\}$ is chosen, one may use the minimization of the functional

$$Z(r,\delta) = \int_{t_1}^{t_2} \left( \frac{d}{dt} \sigma(t,r,\delta) \right)^2 dt, \qquad (2.7)$$

where $r, \delta$ are the parametrization parameters for the whole family of the squeezed states. In a non-steady case, the vacuum state is a conditional notion for the in-vacuum $t_1 \to -\infty$ and the out-vacuum $t_2 \to \infty$ states. As may see, the nonlinear equation appears even in a linear quantum problem for determining a true vacuum state of the time-dependent oscillator.

The examples of the $\sigma(t)$ - behavior for the in- and the out- vacuum states are shown in Fig. 1.

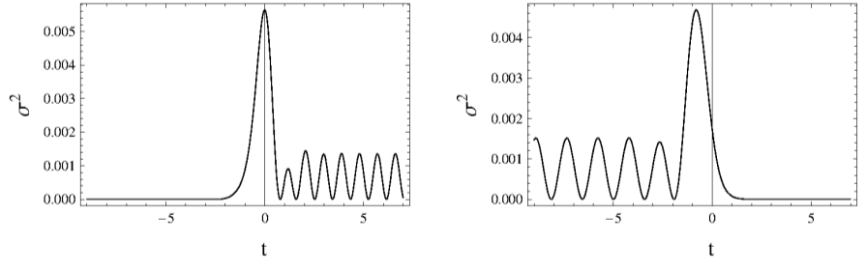

Fig. 1. The examples of the $\sigma^2$ -function behavior for the in- and the out- vacuum states.

Other insight bridging the quantum and classical systems could regard to a decreasing of the dispersion of the dynamical variables mean values. An example is the cosmological mini-superspace model. The Hamiltonian in this model is simultaneously a constraint condition $H = 0$ which should be satisfied alongside with the equations of motion.

Let us consider the toy model with a massless scalar field $\phi$ and the "by hand" introduced decrease of the cosmological constant $V_0$ [3]. The Hamiltonian of the model has the form:

$$H = -\frac{p_a^2}{2a} + \frac{p_\phi^2}{2a^3} + V_0 \frac{a^3}{1+\beta a^3}, \qquad (2.8)$$

where $p_a$ and $p_\phi$ are the momentums associated with the Universe scale-factor $a$ and the scalar field $\phi$, respective, and $\beta$ is some constant. This Hamiltonian assumes a modification of the gravity theory with a cosmological constant in a sense that this "constant" $V_0 \frac{a^3}{1+\beta a^3}$ is non-zero at the small-scale factors and decreases as $\propto a^{-3}$ at the large scale-factors (i.e., it is a model of the terminating inflation).

The corresponding equations of motion are:

$$\ddot{\alpha}+\frac{3}{2}\left(\dot{\alpha}^2+\dot{\phi}^2\right)-\frac{3V_0}{\left(1+\beta e^{3\alpha}\right)^2}=0, \quad \ddot{\phi}+3\dot{\alpha}\dot{\phi}=0, \tag{2.9}$$

where $\alpha \equiv \ln a$.

For quantization, one should consider the Hamiltonian constraint as a condition for a state vector $|\Psi\rangle$: $\hat{H}|\Psi\rangle = 0$. As a result, we come to the Wheeler-DeWitt equation [3,4]:

$$\left(\frac{1}{2a^2}\frac{\partial}{\partial a}a\frac{\partial}{\partial a}-\frac{1}{2a^3}\frac{\partial^2}{\partial \phi^2}+V_0\frac{a^3}{1+\beta a^3}\right)\Psi(a,\phi)=0. \tag{2.10}$$

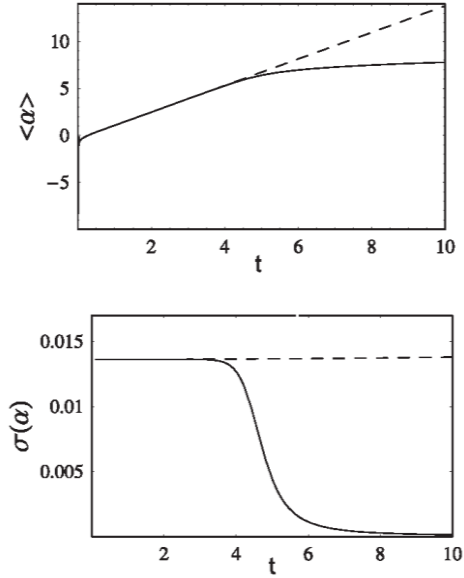

Fig. 2. The mean value of the logarithm of the scale factor $\langle\alpha\rangle$ and its dispersion $\sigma(\alpha)$ for the model (2.8) with the cosmological constant $V_0=1$, $\beta=0$ (dashed curves), and with the decreasing cosmological constant $V_0=1$, $\beta=10^{-8}$ (solid curves).

The paradox is that there is no explicit time-variable in this equation, which manifests the so-called "problem of time" in the quantum cosmology [5]. Formally, the Hamiltonian is the field equation constraint in the general theory of relativity. That means that the total energy of the gravitational field and the matter vanishes. Thus, all states form the Hamiltonian "null-space" after canonical quantization (that results in the Wheeler-DeWitt equation (2.10)). That is all quantum states are "vacuum states" (the Hamiltonian "annihilates" them). But the Hamiltonian provides a time-evolution. Thus, there is no time-evolution in the quantum cosmology.

However, this is rather a pseudo-problem, since the time-evolution remains in the equations of motion (2.9) so that one could only write "hats" over $\hat\alpha$ and $\hat\phi$ to consider them as the quasi-Heisenberg operators and Eqs. (2.9) as the operator equations [6]. The commutation rules for these operators follow from the Dirac brackets for a constraint system. They can be evaluated explicitly at the initial moment of time then the system allows evolving in accordance with the equations of motion. The Hilbert space for the quasi-Heisenberg operators is built on the basis of an asymptotical solution of the Wheeler-DeWitt equation (2.10).

The results of the calculation are shown in Fig. 2. One can see that the Universe becomes "classical" after the inflation end. It means that the sufficiently quick decrease of the cosmological constant causes suppressing the dispersion of the scale factor logarithm.

## 3 The relation with the solitonic and statistical physics

Here we invent the connection with the solitonic physics [7] based on the idea that the classical states are the result of the quantum evolution of a nonlinear system evolving to the state with the small dispersions of the mean values of observables. Thus, the nonlinear equations arise in a quantum linear physics when one tries constructing a vacuum state. On the other hand, one may see that a quantum system tends to classical one in some cases. Thereby, the solitonic physics can be incorporated into the field of quantum physics including both linear and nonlinear phenomenon.

More specifically, a soliton can be interpreted as a coherent structure formed in the self-interacting bosonic system, i.e., as a classical analog of the Bose-Einstein condensate [8-10]. Such a coherent condensate is defined by the two-point correlation function in the momentum $p-$space: $\left\langle A_p(t) A_{p'}^*(t) \right\rangle = n_p \delta(p-p')$, where $A_p(t) = \frac{1}{\sqrt{2\pi}} \int \psi(t,x) e^{-ipx} dx$, $\psi(t,x)$ is a field amplitude, and $n_p$ is a "particle number" distribution characterizing the soliton "shape." The "condensation" means a flow of energy to zero wavenumbers $p \to 0$ that is the increase of long-range correlations and the suppression of fluctuations in direct analogy with minimization of dispersion of a quantum system transiting to a classical state (Fig. 2) [11]. Simultaneously, that results in the minimization of the Hamiltonian $H(\psi)$ defined as

$$H(\psi) = \int \left( \left|\frac{\partial \psi}{\partial x}\right|^2 \mp |\psi|^4 \right) dx \qquad (3.1)$$

for the well-known (1+1)-dimensional cubic nonlinear Schrödinger equation which describes an evolution of slowly varying wave in a nonlinear medium [9,12]:

$$i\frac{\partial \psi}{\partial t} = \frac{\delta H}{\delta \psi^*} \quad \text{or} \quad i\frac{\partial \psi}{\partial t} + \frac{\partial^2 \psi}{\partial x^2} \pm |\psi|^2 \psi = 0. \tag{3.2}$$

Such a coherent condensate (i.e., a soliton) minimizing the Hamiltonian and existing as a steady "ground" state (i.e., $\psi(t,x) = \phi(x)e^{-i\lambda t}$) allows treating as an analog of the vacuum state of the nonlinear system far from the thermodynamic equilibrium.

A soliton (i.e., a coherent "condensate") has the minimal entropy so that the rest of entropy concentrates in the small-scale fluctuations with large $|\psi_x|^2$ outside the condensate [8,13]. As a result, the condensate evolves toward the Rayleigh-Jeans equilibrium distribution [11]:

$$n_p \propto \frac{1}{p^2 - \mu} \Theta\left(p_{cut}^2 - p^2\right) \tag{3.3}$$

which obeys two correlation scales: a long-range one defined by a negative "chemical potential" $\mu$, and a short-range one defined by a momentum cut-off at $p_{cut}$ which is caused by the nonlinear and dissipative effects ($\Theta$ is the Heaviside function, see Fig. 3).

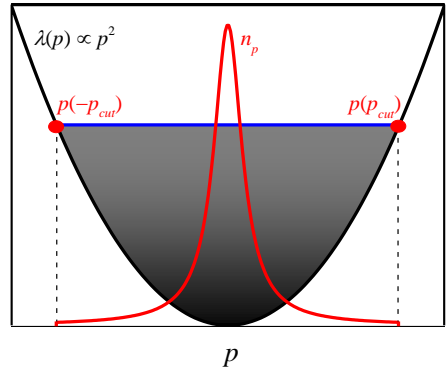

Fig. 3. The Langmuir dispersion relation $\lambda \propto p^2$ (black curve) and the Rayleigh-Jeans equilibrium distribution for a DS or the turbulence (red curve). The condensation in the vicinity of $p = 0$ is illustrated by shading.

Thus, a bridge to the statistical mechanics is in the offing, and such invention is relevant to the description not only coherent solitons but also to the study of the dissipative solitons (DS) and the turbulent phenomena [11,14].

# 4 Phase-space representation of nonlinear dynamics

As was demonstrated above, the phase-space (Hamiltonian) description of nonlinear dynamical systems in both quantum and classical mechanics provides with a guideline in the solution and interpretation of the complex problems that entwines the seemingly disjointed concepts ranging from quantum cosmology to solitonics and statistical mechanics.

Regarding the quantum mechanics operating in a linear operator Hilbert space, we need associating the operator $\hat{A}$ in the $q$-representation with an appropriate function in the Weyl's, Wigner's, Moyal's, and Groenewold's style [15]:

$$\tilde{A}(x,p) = \int e^{-ipq/\hbar} \left\langle x+\frac{q}{2}\left|\hat{A}\right|x-\frac{q}{2}\right\rangle dq, \tag{4.1}$$

and to relate the quantum density with the so-called Wigner function $W(x,p)$ which has a direct association with the probability density operator $\hat{\rho} = |\psi\rangle\langle\psi|$ [16]:

$$W(x,p) \propto \int e^{-ipq/\hbar} \psi\left(x+\frac{q}{2}\right)\psi^*\left(x-\frac{q}{2}\right) dq \tag{4.2}$$

that provides the measurable expectation value of a $\hat{A}$-operator:

$$\langle A \rangle = \iint W(x,p)\tilde{A}(x,p) dx dp. \tag{4.3}$$

Finally, we have to associate the noncommutativity of operators with some ordering rule, e.g., in the Weyl's style:

$$\hat{p}^2 \hat{x} \to \frac{1}{3}\left(\tilde{p}^2\tilde{x} + \tilde{p}\tilde{x}\tilde{p} + \tilde{x}\tilde{p}^2\right). \tag{4.4}$$

Returning to nonlinear optics, the nonlinear Schrödinger equation (3.2) with a potential $U(x)$ allows the phase-space representation through the Wigner transformation

$$W(x,p) \propto \int e^{-ipq} \psi\left(x+\frac{q}{2}\right)\psi^*\left(x-\frac{q}{2}\right) dq \tag{4.5}$$

resulting in [17]:

$$\frac{\partial W(x,p)}{\partial t} \pm p\frac{\partial W(x,p)}{\partial x} + \\ + \sum_{s=0}^{\infty} \frac{(-1)^s}{(2s+1)!2^{2s}} \frac{\partial^{2s+1} U(x)}{\partial x^{2s+1}} \frac{\partial^{2s+1} W(x,p)}{\partial p^{2s+1}} = 0, \tag{4.6}$$

where the self-interaction potential is $U(x) = \int W(x,p) dp$.

Two problems are that the resulting equation (4.6) contains the infinite expansion term, and it is the integrodifferential equation in (2+1)-dimensions. Nevertheless, our calculations demonstrated that the geometrical optics approximation $\Delta x \Delta p \gg 1$ is well-working even for the "true vacuum" (not only

"squeezed one," see below) states and can be modeled by the Vlasov's equation ($s = 0$ in Eq. (4.6)) [18,19]:

$$\frac{\partial W(x,p)}{\partial t} \pm p\frac{\partial W(x,p)}{\partial x} + \frac{\partial U(x)}{\partial x}\frac{\partial W(x,p)}{\partial p} = 0, \quad (4.7)$$

re-interpreting the Wigner function as a probability distribution function:

$$W(x,p) \propto \int e^{-ipq} \left\langle \psi\left(x+\frac{q}{2}\right)\psi^*\left(x-\frac{q}{2}\right)\right\rangle dq \quad (4.8)$$

where $\langle \bullet \rangle$ denotes a statistical average. Eq. (4.7) describes the quasi-particles statistics in the effective self-consistent potential, and we may interpret a soliton as a self-organized ensemble of interacting quasi-particles ("internal modes") and use the methods of statistical mechanics.

Figs. 4-7 demonstrate the evolution of a Wigner function in the so-called anomalous dispersion regime $\beta > 0$, where the classical soliton exists [20]:

$$i\frac{\partial \psi}{\partial t} + \beta\frac{\partial^2 \psi}{\partial x^2} + |\psi|^2 \psi = 0. \quad (4.9)$$

When the nonlinearity prevails over the dispersion, the initial pulse inevitable collapses (Fig. 4). But a compensation of pulse squeezing due to nonlinearity by dispersion results in the soliton formation, which is stable, perfectly localized and coherent structure (Fig. 5). When the dispersion prevails over nonlinearity, the pulse spreads in the time domain but with the conservation of its spectral width. It is an example of squeezing described by the so-called "chirp" parameter $\theta$, that is a slope of the Wigner function in our case (Figs. 6,7).

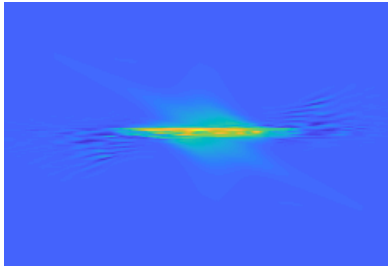

Fig. 4. The Wigner function for the dimensionless dispersion $\beta = 1$ after the 6 dimensionless nonlinear propagation lengths.

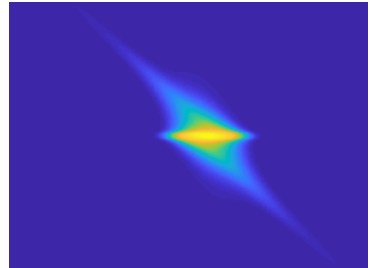

Fig. 5. The Wigner function for the dimensionless dispersion $\beta = 2$.

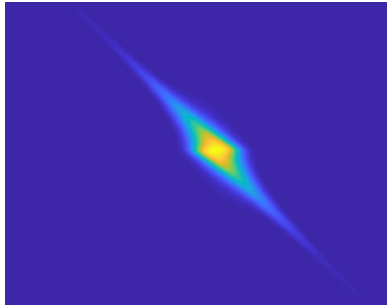 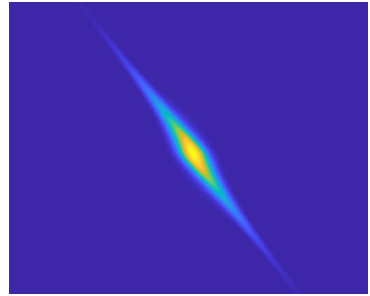

Fig. 6. The Wigner function for the dimensionless dispersion $\beta = 3$.

Fig. 7. The Wigner function for the dimensionless dispersion $\beta = 4$.

In the normal dispersion regime, the tendency to collapse is arrested, so that the energy concentration at zero wave-number (carrier frequency) results in a squeezing state with a huge "chirp" (Fig. 8)

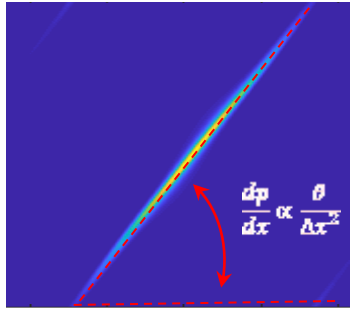

Fig. 8. The Wigner function for the dimensionless dispersion $\beta = -4$.

However, this state is not steady. It tends to disappear in the "fluctuation sea." One may propose a way out of this problem: let's make our system dissipative that could provide an inverse energy cascade outward the zero wave-number but without the coherency loss. The example of such open system is a laser with the linear and nonlinear gain, loss ($\sigma, \kappa$), and the spectral dissipation ($\alpha$).

$$i\frac{\partial \psi}{\partial t} + \beta \frac{\partial^2 \psi}{\partial x^2} + |\psi|^2 \psi = i\left(\sigma \psi + \alpha \frac{\partial^2 \psi}{\partial x^2} + \kappa |\psi|^2 \psi\right). \qquad (4.10)$$

The combination of these factors provides a right energy redistribution $E = 2\left(\sigma |\psi|^2 + \kappa |\psi|^4 - \alpha \left|\frac{\partial \psi}{\partial x}\right|\right) + \alpha \frac{\partial^2 |\psi|^2}{\partial x^2}$ (Fig. 9) that stabilizes the DS coherent structure.

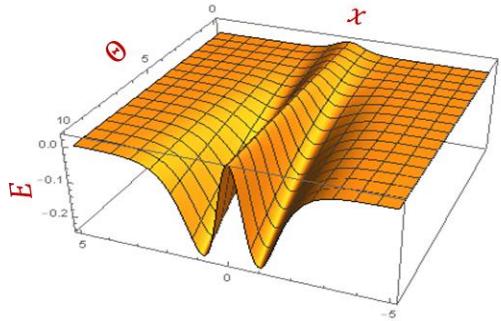

Fig. 9. The energy flow $E$ inside of a DS.

Here, we deal with a DS with a nontrivial internal structure providing a huge chirp without spectral squeezing that allows a coherent energy harvesting. This huge chirp validates the lowest-order term approximation in the Weyl-Wigner equation (4.7) and we reveal with surprise that a dissipative soliton is a self-organized "ensemble" of self-interacting quasi-particles, somewhat like an elementary "community," and the methods of statistical mechanics could allow the description of such metaphorical "community" without a direct statistic gathering of the "individual fates."

## Conclusions

Phase-space formulation of non-stationary nonlinear systems reveals an affinity between seemly different branches of physics such as dynamics of nonlinear systems far from equilibrium, statistical mechanics, thermodynamics, and quantum physics. One of the key insights is a clear distinction between the "vacuum" and "squeezed" states of a system. A soliton can be treated as a "vacuum state" of a closed nonlinear system, and such low-entropy state minimizes a Hamiltonian so that the second law of thermodynamics needs an entropy concentration in small-scaled (down to quantum level) fluctuations. The "squeezed states" (or coherent "condensates") mimic vacuum ones and can be very attractable in praxis, in particular, for energy harvesting at ultrashort time scales. However, such states are not steady-state in a closed system. The stabilization of such coherent structure is possible in an open, i.e., dissipative system. That means a DS formation. The phase-space analysis demonstrates a close analogy between DS and turbulence phenomena in plasma and condensed media that allows formulating the statistical mechanics and quantum approaches to the extremely broad diapason of nonlinear phenomena. In particular, the promising advantage of the phase-space formulation of the DS dynamics is the possibility of direct calculation of statistical (including quantum) properties coherent, partially-coherent, and non-coherent dissipative structure without numerically consuming statistic harvesting.